\documentclass[letterpaper, 10 pt, conference]{ieeeconf}  

\IEEEoverridecommandlockouts                              

\usepackage{amsfonts}
\usepackage{mathrsfs}
\usepackage{amssymb}
\usepackage{extarrows}
\usepackage{color}
\usepackage{cite}

\def\rep#1{(\ref{#1})}

\newcommand{\R}{\mathbb{R}}

\def\C{\;{\rm l\!\!\!C}}

\def\send#1#2{\stackrel{#1}{\hbox to #2{\rightarrowfill}}}
\def\-{\!\!\!\!\!-}

 \def\qed{ \rule{.1in}{.1in}}
\def\eq#1{\begin{equation}#1\end{equation}}

\newcommand{\rank}{{\rm rank\;}}

\def\scr#1{{\cal #1}}

\newcommand{\matt}[1]{\begin{bmatrix}#1\end{bmatrix}}

\newcommand{\dfb}{\stackrel{\Delta}{=}}

\newtheorem{theorem}{Theorem}
\newtheorem{lemma}{Lemma}

\newtheorem{proposition}{Proposition}

\def\e{{\bf e_T}}

\def\qed{ \rule{.1in}{.1in}}

\def\R{{\rm I\!R}} 

\newcounter{seqn}[equation]
\def\theseqn{\arabic{equation}\alph{seqn}}

\def\endseqn{\eqno \@seqnnum
$$\ignorespaces}
\def\@seqnnum{(\theseqn)}

\newskip\mcentering \mcentering=0pt plus 1000pt minus 1000pt

\def\meqalignno#1{
\halign to\displaywidth{
    \hbox to 0pt{\kern\displaywidth\llap{$##$}\hss}\tabskip=\mcentering
    &\hfil$\displaystyle{##}$\tabskip=\mcentering
   &&$\displaystyle{{}##}$\hfil\tabskip=\mcentering
    \crcr
    #1\crcr}}

\def\rep#1{(\ref{#1})}

\def\eq#1{\begin{equation}#1\end{equation}}

\def\dspace{\multiply\normalbaselineskip 150
		  \divide\normalbaselineskip 100 \normalbaselines
		  \csname @@normalbaselineskip\endcsname\normalbaselineskip}
\def\sspace{\multiply\normalbaselineskip 200
		 \divide\normalbaselineskip 300 \normalbaselines
		 \csname @@normalbaselineskip\endcsname\normalbaselineskip}
\def\sdspace{\multiply\normalbaselineskip 160
		 \divide\normalbaselineskip 150 \normalbaselines
		 \csname @@normalbaselineskip\endcsname\normalbaselineskip}

\def\@{\tilde}

\def\3dot#1{\buildrel\textstyle...\over#1}

\title{\LARGE \bf
Distributed Control of Linear Multi-Channel Systems}

\author{L. Wang, D. Fullmer, F. Liu, and A. S. Morse
\thanks{A full-length version of this paper including proofs is currently posted on the archive (http://arxiv.org/abs/1909.11823) and will be submitted for journal publication in the near future.}
\thanks{L. Wang, D. Fullmer, F. Liu, and A. S. Morse are with the Department
of Electrical Engineering, Yale University
{\tt\small \{lili.wang, daniel.fullmer, fengjiao.liu,  as.morse@yale.edu\}}.
}%
\thanks{This research was supported by  AFOSR, ARO, NSF}}

\begin{document}

\maketitle

\thispagestyle{empty}

\begin{abstract}A solution is given to the basic distributed  feedback control problem for a multi-channel linear system assuming
 only that the system is jointly controllable, jointly observable
and has an associated neighbor graph which is strongly connected. The solution is an  observer-based  control system which is implemented
 in a distributed manner. Using these ideas, a solution is also given to the distributed set-point control problem for a multi-channel linear system
  in which each and every agent with access to the system is able to independently adjust its controlled output to any desired set-point value.
  An example is given to briefly illustrate how network transmission delays
  might be dealt with.
  \end{abstract}

\section{Introduction}

A well-known  application of an observer
    is to serve as a component
  of a feedback control system for  regulating a dynamical process. In particular,
  for a  controllable and observable linear system  $y=Cx$, $\dot{x} = Ax +Bu$, such a feedback control
   is  of the form $u = F\hat{x}$ where $\hat{x}$ is the state of the observer $\dot{\hat{x}} =
(A+KC)\hat{x}-Ky+F\hat{x}$, and $F$ and $K$ are matrices  which are usually chosen so that
 $A+BF$ and $A+KC$ are at least stability matrices. As is also well known, the resulting closed-loop system
  can be   described
 by the equations
 \begin{eqnarray} \dot{x} &=& (A+BF)x +BFe\\\dot{e} &=& (A+KC)e\label{e1}\end{eqnarray}
 where $e$ is the state estimation error $e=\hat{x}-x$. As is plainly clear,  the utility of this approach is in no small part due to the fact that
   the error
  dynamics described by \rep{e1} is an {\em unforced} linear differential equation. Arriving at this necessitates including in the
  differential equation defining the observer, the term $F\hat{x}$. While this is a perfectly valid step for the
  centralized observer under discussion, an analogous step for the observer-based distributed control of a multi-channel linear system usually cannot be
  carried without violating distributional requirements. The primary aim of this paper is to explain how to overcome this
   difficulty and in so doing, to provide what is almost certainly the first systematic procedure
    for constructing a distributed feedback control for stabilizing or otherwise regulating
      a  multi-channel linear system.


\section{The Problem}\label{p}

Perhaps  the most basic  problem in distributed feedback control  is to develop a procedure which can enable
 a networked family of $m>1$ agents
 to stabilize
or otherwise control in a distributed manner, a physical process $\mathbb{P}$ modelled by a multi-channel,
 time-invariant, linear system.
By an $n$-dimensional,  $m$-{\em channel linear system} is 
  meant a linear system of the form
\eq{\dot{x} = Ax +\sum_{i=1}^mB_iu_i\hspace{0.2in} y_i = C_ix, \;\;i\in\mathbf{m}\label{mcs}}
where, $n$ and $m$ are positive integers,  $\mathbf{m} = \{1,2,\ldots, m\}$,  $x\in\R^n$, and for $i\in\mathbf{m}$,
 $u_i\in\R^{m_i}$ and $y_i\in\R^{p_i}$ are  the control input to channel $i$ and the measured output from channel $i$ respectively.
  Here   $A$, the $B_i$, and the $C_i$ are real-valued,
  constant matrices of appropriate sizes. Without  any real loss of generality it is assumed that the system defined by
  \rep{mcs} is both
   jointly controllable and jointly observable; that is, the matrix pairs
   $$(A,\matt{B_1 &B_2&\ldots &B_m})\;\;\;\; \text{and}\ \;\;\;\left (\matt{C_1 \\C_2\\\vdots \\C_m},A\right )$$
   are controllable and observable respectively. For simplicity it is further assumed
  that $B_i\neq 0,C_i\neq 0,\;\;i\in\mathbf{m}$.

  It is presumed that the system described by \rep{mcs} is to be controlled by $m$ agents with
   the understanding that for $i\in\mathbf{m}$, agent $i$  can measure the $y_i$ and  has access to the  control input $u_i$.
In addition, each agent $i$ can  receive information from its ``neighbors'' where exactly who each agent's neighbors
  are is specified in the problem formulation.
   In this paper it is assumed that  each agent's neighbors do not change with time,
    that $\scr{N}_i\dfb \{j_1^i,j_2^i,\ldots,j_{k_i}^i\}$ is the set of labels of agent $i$'s neighbors including itself,
    and that each agent
  can receive the current state of each of its neighbor's  controllers. Neighbor relations
  can be conveniently described by a directed graph $\mathbb{N}$ defined on $m$ vertices with a
  direct arc from vertex $j$ to vertex $i$
just in case agent $j$ is a neighbor of agent $i$. It is assumed
throughout this paper that $\mathbb{N}$ is strongly connected. It is straightforward to extend what follows
 to the  general  case
 when $\mathbb{N}$ is only weakly connected.

The  basic distributed control problem for the $m$ channel system \rep{mcs}  is to develop a systematic procedure for constructing
$m$ linear time-invariant feedback controls, one for each channel, so that the state of the resulting closed-loop system
converges to zero exponentially fast at a pre-assigned rate.
 Before addressing this problem, it will be useful to briefly review the main results from classical decentralized control \cite{wangdavison,corfmat}.

\subsection{Decentralized Control}

The classical decentralized control problem  for an $m$-channel linear system  is exactly the same as
the distributed control problem just formulated, except for one important difference. In the case of decentralized control, there is no communication between agents
so the only signal available to each agent $i$ is $y_i$.
 The fundamental decentralized control question is this. Under what conditions do there exist
 local linear, time-invariant   controllers, one for each channel, which stabilize  $\mathbb{P}$?
  In answering this question it was shown in \cite{wangdavison}
  that no matter what the local controllers are, the spectrum of the  resulting  closed-loop system contains
   a uniquely determined subset of eigenvalues which remain unchanged no matter which local controllers are applied. This
  is  the {\em fixed spectrum}\footnote{Referred to as ``fixed modes'' in \cite{wangdavison}} of $\mathbb{P}$ \cite{wangdavison,clements}. Decentralized stabilization
   of $\mathbb{P}$ by time invariant linear controls thus  demands that its fixed spectrum  contain only
 open  left half plane eigenvalues. This condition on the fixed spectrum of $\mathbb{P}$
  is necessary and  sufficient for stabilization with decentralized control \cite{wangdavison}.
 In addition,  it is known that that the necessary and sufficient condition for the closed-loop spectrum
 to be freely assignable with decentralized control
 is that $\mathbb{P}$ has no fixed eigenvalues \cite{corfmat}.

The preceding prompts the following question.  Does the distributed control problem formulated  at the beginning of this section have a fixed-spectrum
 constraint analogous to the fixed spectrum constraint encountered in the decentralized control problem?
 The findings of this paper establish  that it does not. This will be accomplished by explaining how to construct a distributed observer-based control system
 which solves the distributed spectrum assignment problem for the multi-channel system described by \rep{mcs}. We begin with a brief review of distributed observers.

\section{Distributed Observer}\label{di}

In a series of papers \cite{KhanAli2010CDC,martins,TAC.17,MitraPurdue2016,Kim2016CDC,trent,ACC19,CDC19.1,TAC.hybrid,sanfelice}, a variety of
distributed observers have
 been proposed for estimating the state of \rep{mcs} assuming
 all of the $u_i=0$. The distributed observer studied in \cite{TAC.17} will be used in this paper. It  is described by the equations
 \begin{eqnarray} \dot{x}_i\!\!\!\! &=&\!\!\!\! (A+K_iC_i)x_i -K_iy_i +
 \sum_{j\in\scr{N}_i}H_{ij}(x_i-x_j)
 \nonumber \\ &\ &
 +\delta_{iq}\bar{C}z,\;\;\;i\in\mathbf{m}\label{mm1}\\
\dot{z} \!\!\!\!& = & \!\!\!\!\bar{A}z  +\bar{K}C_qx_q -\bar{K} y_q + \sum_{j\in\scr{N}_q}\bar{H}_{j}(x_q-x_j) \label{mm2} \end{eqnarray}
\noindent where all $x_i\in\R^n$, $z\in\R^{m-1}$,  $q\in \mathbf{m}$,
 $\delta_{iq}$ is the Kronecker  delta, and the $K_i,H_{ij},\bar{A},\bar{K}, \bar{H}_j,\bar{C}$ are matrices of appropriate sizes. The subsystem consisting of
 \rep{mm2} and the signal $\delta_{iq}\bar{C}z$ is called a {\em channel controller } of \rep{mm1}.  Its function will be explained in the sequel.

The error dynamics for this observer are described by the equations
\begin{eqnarray}\dot{e}_i\!\!\! &= &\!\!\!\!(\!A+\!K_iC_i)e_i + \!\!\!\sum_{j\in\scr{N}_i}\!\!\!H_{ij}(e_i\! -\!e_j) +\delta_{iq}\bar{C}z\label{xe1}\\
\dot{z}\!\!\! & = &\!\!\!\! \bar{A}z  +\bar{K}C_qe_q + \sum_{j\in\scr{N}_q}\bar{H}_{j}(e_q-e_j)\label{xe2}  \end{eqnarray}
where for $i\in\mathbf{m}$, $e_i$ is the $i$th state estimation error $e_i= x_i-x$.
Note that \rep{xe1}, \rep{xe2}
is an $(mn +m-1)$-dimensional, unforced linear system.
 It is known that
its spectrum  can be freely assigned  by appropriately picking the matrices
$K_i,H_{ij},\bar{A},\bar{K}, \bar{H}_j,\bar{C}$ \cite{TAC.17}. Thus by so choosing these matrices, all of the $e_i$ and $z$ can be made to converge
 to zero exponentially fast at a pre-assigned rate.

There are several steps involved in picking these matrices. First $q$ is chosen; any value of $q\in\mathbf{m}$ suffices.
 The next step is to temporarily ignore the channel controller \rep{xe2}
and to choose matrices $\tilde{K}_i$ and the $\tilde{H}_{ij}$ so that the {\em open-loop error system}
\eq{\dot{e}_i = (A+\tilde{K}_iC_i)e_i + \sum_{j\in\scr{N}_i}\tilde{H}_{ij}(e_i-e_j) +\delta_{iq}\tilde{u}_q,\label{ppe1}
} $i\in\mathbf{m}$, is controllable by $\tilde{u}_q$ and observable through
\eq{\tilde{y}_q = \matt{C_qe_q \\ e_q-e_{j^q_1} \\ \vdots \\e_q-e_{j^q_{k_q}}}\label{ppe2}}
where $ \{j_1^q,j_2^q,\ldots,j_{k_q}^q\} = \scr{N}_q$.
In fact, the set of $\tilde{K}_i, \tilde{H}_{ij},\;j\in\scr{N}_i$
 for which these properties hold is the complement of a proper algebraic set in the linear space of all such matrices \cite{TAC.17}.
 Thus almost any choice for these matrices will accomplish the desired objective.

 The next step is to pick  matrices $\bar{A}$, $\bar{B}$, $\bar{C}$ and $\bar{D}$ so that
 so that the closed loop spectrum of the system consisting of \rep{ppe1}, \rep{ppe2} and the channel controller
$$ \tilde{u}_q = \bar{C}z
 +\bar{D}\tilde{y}_q,\;\;\;\;\;  \dot{z} = \bar{A}z +
 \bar{B}\tilde{y}_q$$ has the prescribed spectrum. One technique for choosing these matrices can be found in \cite{braschpearson}.
 Of course since the
 system defined by  \rep{ppe1} and \rep{ppe2} is controllable and observable,  there are many ways to define  a channel controller and thus the
  matrices $\bar{A},\bar{B},\bar{C},$ and $\bar{D}$.  In any event, once these matrices are chosen,
the $K_i$ and $H_{ij}$ are defined so that for all $i\neq q$ , $K_i \dfb \tilde{K}_i$ and $H_{ij} \dfb \tilde{H}_{ij}, j\in\scr{N}_q$,  while for $i=q$,
 $K_q \dfb \tilde{K}_q+\hat{K}_q$ and $H_{qj} \dfb \tilde{H}_{qj} + \hat{H}_{qj},\;j\in\scr{N}_q,$ where
$ \matt{\hat{K}_q &\hat{H}_{qj^q_1} &\cdots  \hat{H}_{qj^q_{k_q}}} = \bar{D}$. Finally $\bar{K}_q$ and the $\bar{H}_{j},\;j\in\scr{N}_q$ are defined so that
$\matt{\bar{K}_q &\bar{H}_{j^q_1} &\cdots  \bar{H}_{j^q_{k_q}}} = \bar{B}$.

\section{Distributed-Observer Based Control}

 The first step in the development
 of a distributed observer based feedback system for \rep{mcs} is to devise state feedback laws $u_i = F_ix,\;i\in\mathbf{m}$,
 which endow the closed loop system
\eq{\dot{x} = \left (A +\sum_{i=1}^mB_iF_i\right )x \label{ssxmcs}}
with prescribed properties such as stability and/or optimality with respect to some performance index.
In accordance with certainty equivalence,  the next step is to implement instead of state feedback laws   $u_i = F_ix,\;\;i\in\mathbf{m}$, the distributed
 feedback laws $u_i = F_ix_i,\;\;i\in\mathbf{m}$, where $x_i$ is agent $i$'s estimate of $x$ generated by a distributed observer.
 Doing this results in the system
\eq{\dot{x} = Ax +\sum_{i=1}^mB_iF_ix_i \label{2ssxmcs}}
instead of\rep{ssxmcs}.

A system which   provides the required estimates $x_i$ of $x$ is
 \begin{eqnarray}\dot{x}_i\!\!\!\!\! &=&\!\!\! \!\!(A+K_iC_i)x_i -K_iy_i +
 \sum_{j\in\scr{N}_i}H_{ij}(x_i-x_j) \nonumber
 \\ &\ &
 +\delta_{iq}\bar{C}z + \sum_{j=1}^mB_jF_jx_j  \;\;\;i\in\mathbf{m}\label{nnq1}\\
\dot{z}\!\!\!\!\! & = & \!\!\!\!\!\bar{A}z  +\bar{K}C_qx_q -\bar{K} y_q + \!\!\!\sum_{j\in\scr{N}_q}\!\!\!\bar{H}_{j}(x_q-x_j) \label{nnq2}  \end{eqnarray}
since, in this case the associated error system is exactly the same as before when there was no
 feedback to the process to account for.  Unfortunately this system cannot be used without violating the  problem
  assumptions since the implementation of \rep{nnq1}  requires each agent to use the state estimates of those agents
  which are not its neighbors.
An alternative system which is implementable without violating problem assumptions is the modified
 distributed state estimator
\begin{eqnarray}\dot{x}_i\!\!\! &=&\!\!\! (A+K_iC_i)x_i -K_iy_i +
 \sum_{j\in\scr{N}_i}H_{ij}(x_i-x_j)
 \nonumber \\  &\ &
 +\delta_{iq}\bar{C}z + \left (\sum_{j=1}^mB_jF_j\right )x_i,  \;\;\;i\in\mathbf{m}\label{me1}\\
\dot{z}\!\!\! & = &\!\!\! \bar{A}z  +\bar{K}C_qx_q -\bar{K} y_q + \!\!\!\sum_{j\in\scr{N}_q}\!\!\!\bar{H}_{j}(x_q-x_j)\label{me2}   \end{eqnarray}
In the sequel it will be shown that even with this modification, this system can still provide the required estimates of $x$.

The error dynamics for \rep{me1}, \rep{me2} are described by  the  linear system
 \begin{eqnarray}\dot{e}_i &=& (A+K_iC_i)e_i  +
 \sum_{j\in\scr{N}_i}H_{ij}(e_i-e_j)
 +\delta_{iq}\bar{C}z
 \nonumber \\ & \ &
 + \sum_{j=1}^mB_jF_j(e_i-e_j),  \;\;i\in\mathbf{m}\label{eme1}\\
\dot{z} & = & \bar{A}z  +\bar{K}C_qe_q  + \sum_{j\in\scr{N}_q}\bar{H}_{j}(e_q-e_j) \label{eme2}  \end{eqnarray}
while the process dynamics modelled by \rep{2ssxmcs}  can be rewritten   as
\eq{\dot{x} = \left (A +\sum_{i=1}^mB_iF_i\right )x +\sum_{i=1}^mB_iF_ie_i \label{3ssxmcs}}
Since \rep{eme1}, \rep{eme2} is an unforced linear system, its dynamic behavior is determined primarily by its spectrum.
In the sequel it will be  explained how to choose the $K_i$, $H_{ij}$, $\bar{K}_i$ and $\bar{H}_i$  so that the spectrum of
\rep{eme1} and \rep{eme2} coincides with a prescribed   symmetric set
 of complex numbers.
To achieve this, attention will first be focused on the properties of the  open-loop error system    described by
\begin{eqnarray}
\dot{e}_i &= & (A+K_iC_i)e_i  +
 \sum_{j\in\scr{N}_i}H_{ij}(e_i-e_j)
 \nonumber \\ &\ &+ \sum_{j=1}^mB_jF_j(e_i-e_j) +\delta_{iq}\tilde{u}_q, \;\;i\in\mathbf{m}  \label{auditt}\end{eqnarray}
and \rep{ppe2}. This system is what what results when the  channel controller appearing in \rep{eme2} is removed.
The main technical result of this paper is as follows.

\begin{proposition} There are matrices $K_i,  H_{ij},j\in\scr{N}_i, \;i\in\mathbf{m} $ such that for all $q\in\mathbf{m}$,
 the open-loop error system described by \rep{auditt} and \rep{ppe2}     is observable through $\tilde{y}_q$ and  controllable
 by $\tilde{u}_q$  with controllability index $m$.
\label{maint}\end{proposition}

\noindent The implication of this proposition is clear.

\begin{theorem} For any set of feedback matrices $F_i,i\in\mathbf{m}$,  any integer $q\in\mathbf{m}$, and any symmetric set of
 $mn + m-1$ complex numbers $\Lambda$,  there are matrices $K_i,\bar{K}_i,H_{ij},\bar{H}_i$ for which the spectrum of the closed-loop error system defined by \rep{eme1} and \rep{eme2}
  is $\Lambda$. \label{main}\end{theorem}

We will now proceed to justify Proposition \ref{maint}.
First  note that \rep{auditt} can be written in the compact form
\eq{\dot{\epsilon} = \left (\tilde {A}  + \sum_{i=1}^m \tilde{B}_i(K_i\hat{C}_i+H_{i}\tilde{C}_{i})\right
 )\epsilon+\tilde{B}_q\tilde{u}_q\label{com}}
\noindent where $\epsilon = \text{column}\{e_1,e_2,\ldots, e_m\}$. Here $\tilde{A} = I_{m\times m}\otimes(A+\sum_{j=1}^mB_jF_j) -Q$ where
$\otimes $ is the Kronecker product and
$Q$ is the $nm\times nm$ partitioned matrix of $m^2$ square blocks, whose $ij$th block
 is  $B_jF_j$;
$\tilde{B}_i$ is the matrix  $\tilde{B}_i = b_i\otimes I_{n\times n},\;i\in\mathbf{m}$ where $b_i$ is
the $i$th unit vector in $\R^m$ and $\hat{C}_i = C_i\tilde{B}_i'$;
$H_i$ is the matrix
   $H_i = \matt{H_{ij^{i}_1} & H_{ij^i_2}&\cdots& H_{ij^i_{k_{i}}}}$ where
 $\{j^i_1,j^i_2,\ldots, j_{k_i}^i\}=\scr{N}_i$; $\tilde{C}_i$ is the matrix $\tilde{C}_i
  = \text{column}\{C_{ij^i_1}, C_{ij^i_2},\ldots, C_{ij^i_{k_{i}}}\}$ where
 $C_{ij} = c_{ij}\otimes I_{n\times n},\;j\in\scr{N}_i,\;i\in\mathbf{m}$ and
 $c_{ij}$ is the row in the transpose of the incidence matrix
of $\mathbb{N}$ corresponding to the arc  from $j$ to $i$.

Next  observe that \rep{com} is what results when the distributed feedback law
$\tilde{v}_i = \matt{K_i &H_i}\tilde{y}_i +\delta_{iq}\tilde{u}_q, \;i\in\mathbf{m}, $ is applied to the $m$ channel linear system
\eq{ \dot{\epsilon} = \tilde{A}\epsilon +\sum_{j=1}^m\tilde{B}_j\tilde{v}_j,
\;\;\tilde{y}_i = \matt{\hat{C}_i\\\tilde{C}_i}\epsilon,\;\;i\in\mathbf{m}\label{anal}}
The proof of Proposition \ref{maint} depends on the following lemmas.

\begin{lemma} The $m$-channel linear system described by \rep{anal} is jointly controllable and jointly observable.\label{lanal}\end{lemma}

\noindent{\bf Proof of Lemma \ref{lanal}:}
In view of the definitions of the $\tilde{B}_i$ it is clear that
 $\matt{\tilde{B}_1 &\tilde{B}_2 &\cdots &\tilde{B}_m}$ is the $nm\times nm$ identity.
  Therefore that \rep{anal} is jointly controllable.

To establish joint observability, suppose that $\tilde{v}$ is an eigenvector of $\tilde{A}$ for which
$$ \matt{\hat{C}_i\\\tilde{C}_i}\tilde{v}=0,\;\;\;\;i\in\mathbf{m}$$ From the relations
$\tilde{C}_i\tilde{v}=0,\;\;\;\;i\in\mathbf{m}$,
 the definitions of the $\tilde{C}_i$ and the assumption that $\mathbb{N}$ is strongly connected it follows that
 $\tilde{v} =\text{column}\{v,v,\ldots, v\}$ for some vector $v\in\R^n$. Meanwhile from the relations
  $\hat{C}_i\tilde{v}=0,\;\;\;\;i\in\mathbf{m}$ and the definitions of the $\hat{C}_i$ it
   follows that $C_iv = 0,\;\;i\in\mathbf{m}$. Moreover from the definition of $\tilde{A}$ and the structure of $\tilde{v}$
     it is clear that $\tilde{A}\tilde{v} =( I_{m \times m}\otimes A) \tilde{v} = \text{column}\{Av,Av,\ldots,Av\}$.
      This and the hypothesis that $\tilde{v}$ is an eigenvector of $\tilde{A}$ imply that $v$  must be an eigenvector of $A$.
      But this is impossible because of joint observability of \rep{mcs} and the fact that $C_iv = 0,\;\;i\in\mathbf{m}$.
      Thus \rep{anal} has no unobservable modes through the combined outputs
      $\tilde{y}_i,\;i\in\mathbf{m}$ which means that the system is jointly observable. $\qed$

\begin{lemma} For any given set of  appropriately sized matrices $K_i,\;i\in\mathbf{m}$,  there exist matrices $H_i$ for which the matrix pair
$\left (\tilde {A}  + \sum_{i=1}^m \tilde{B}_i(K_i\hat{C}_i+H_{i}\tilde{C}_{i}),\tilde{B}_q\right )$
is controllable with controllability index $m$ for every $q\in\mathbf{m}$.
\label{lem2}\end{lemma}

\noindent The proof depends on Lemma \ref{lanal}  and the following facts.

\begin{lemma}There are matrices
$\hat{H}_i,i\in\mathbf{m}$, for which $\left ( \sum_{i=1}^m \tilde{B}_i\hat{H}_{i}\tilde{C}_{i},\tilde{B}_q\right )$
is a controllable pair with controllability index $m$ for every choice of $q\in\mathbf{m}$.\label{daniel}\end{lemma}
\noindent A proof of this lemma will be given below.



\begin{lemma}\label{G-exists}
    Fix $q \in \mathbf m$ and let $b_q$ denote the $q$th unit vector $\mathbb R^m$.
    There exists a matrix $G = \begin{bmatrix}g_{ij}\end{bmatrix} \in \mathbb R^{m \times m}$ with row sums all equal zero and $g_{ij}=0$ whenever agent $j$ is not a neighbor of agent $i$
    such that $(G, b_q)$ is a controllable pair.
\end{lemma}
\noindent

\noindent{\bf Proof of Lemma \ref{G-exists}:}
    Hautus's lemma\cite{hautus} assures that the pair $(G, b_q)$ is controllable if and only if for each eigenvalue $s$ of $G$,
    \begin{equation}
        \begin{bmatrix}G - s I & b_q \end{bmatrix}
    \end{equation}
    has full rank.
    This is equivalent to showing that if $G'x = s x$, and $b_q' x = 0$ then $x=0$.

    Since $\mathbb N$ is strongly connected, there exists a directed spanning tree of $\mathbb N$ whose root is vertex $q$ with all arcs oriented away from $q$,
    which we denote using $\mathbb T_q$.
    Since $\mathbb T_q$ is a directed tree, each vertex $i \in \mathbf m$ has a set of out-neighbors $\mathcal C_i \subset \mathbf m$, and each vertex $i \neq q$ has a unique in-neighbor $\rho_i \in \mathbf m$.

    Choose $v \in \mathbb R^m$ so that $v_q = 0$ and for each $i \neq q$, $v_i$ is a distinct nonzero real value.
    By ``distinct'', we require that $v_i \neq v_j$ for any $i \neq j$.
    Choose $G$ so that, for each $i,j \in \mathbf m$,
    \begin{equation}\label{G-choice}
        g_{ij} = \begin{cases}
            v_i & \textrm{ if } i=j \\
            -v_i & \textrm{ if } j=\rho_i \\
            0 & \textrm{ otherwise}
        \end{cases}
    \end{equation}
    It is clear that $g_{ij}=0$ if $j$ is not a neighbor of $i$ as $\mathbb T_q$ is a subgraph of $\mathbb N$, and $g_{ii} = -\sum \limits_{j=1, j \neq i}^m g_{ij}$.

    Now, suppose $G' x = s x$ and $b_p' x = 0$.
    From $G'x =s x$ and \rep{G-choice},
    \begin{equation}\label{tree-eqn}
        v_i x_i - \sum_{j \in \mathcal C_i} v_j x_j = s x_i \quad i \in \mathbf m
    \end{equation}
    written alternatively,
    \begin{equation}\label{tree-eqn-2}
        (v_i -s) x_i = \sum_{j \in \mathcal C_i} v_j x_j \quad i \in \mathbf m
    \end{equation}
    Additionally, since $b_q'x=0$,
    \begin{equation}\label{root-0}
        x_q = 0.
    \end{equation}

    Since each $v_i,\ i \in \mathbf{m}$ is distinct, there is at most one $i \in \mathbf m$ with $v_i = s$.
    If there is one, choose $r \in \mathbf{m}$ so that $v_r = s$, otherwise, choose $r \in \mathbf m$ arbitrarily.
    Since $\mathbb T_q$ is a tree there must be a unique path from $q$ to $r$,
    let $\mathcal P$ denote the set of labels of vertices along that path, excluding $r$,
    and for each $i \in \mathcal P$,
    let $c_i \in \mathcal C_i \cap \mathcal P$ denote the unique vertex along this path.

    For each nonnegative real number $d$, let $\mathcal V_d \subset \mathbf m$ consist of the set of vertices of depth $d$ in $\mathbb T_q$.
    By induction, we show that for each $d \ge 0$ and $i \in \mathcal V_d$,
    \begin{equation}\label{inductive-hyp}
        x_i = \begin{cases}
            x_r & \textrm{if $i=r$} \\
            \frac{v_{c_i}}{v_{i} - s} x_{c_i} & \textrm{if } i \in \mathcal P \\
            0 & \textrm{otherwise}
        \end{cases}
    \end{equation}

    Suppose $D$ is the maximum depth of $\mathbb T_q$.
    No vertex $i \in \mathcal V_D$ may have any children $\mathcal C_i$, since otherwise it would not be a vertex of maximum depth.
    If $i=r$ then \rep{inductive-hyp} is clearly true.
    Otherwise $i \neq r$, and from \rep{tree-eqn-2}, $(v_i - s) x_i = 0,\ i \in \mathcal V_D$.
    But then $(v_i - s) \neq 0$ since each $v_i$ is distinct and $v_i \neq v_r =s$, so $v_i = 0$.
    Since $\mathcal V_D \cap \mathcal P = \emptyset$ and $x_i = 0$ for each $i \in \mathcal V_D,\ i \neq r$,
    so \rep{inductive-hyp} must hold for vertices $i \in \mathcal V_D$.

    Next, suppose for some $0 < d < D$, \rep{inductive-hyp} is true for vertices in $\mathcal V_d$.
    Suppose also that $i \in \mathcal V_{d-1}$.
    Again, if $i=r$ then \rep{inductive-hyp} is clearly true.
    Otherwise $i \neq r$.
    Noting that each $j \in \mathcal C_i$ is in $\mathcal V_d$,
    the inductive hypothesis~\rep{inductive-hyp} assures that for any $j \in \mathbf m$ with $j \neq r$ and $j \notin \mathcal P$, $v_j = 0$.
    So, from~\rep{tree-eqn-2},
    \begin{equation}
        (v_i - s) x_i = v_{c_i} x_{c_i}
    \end{equation}
    if $i \in \mathcal P$ and
    \begin{equation}
        (v_i - s) x_i = 0
    \end{equation}
    otherwise.
    Note that $v_i  - s \neq 0$ since only $v_r = s$ and $i \neq r$.
    Thus,~\rep{inductive-hyp} holds for vertices in $\mathcal V_{d-1}$ as well.
    So by induction, it holds for each $d \ge 0$ and $i \in \mathcal V_d$.

    Repeated substitution of~\rep{inductive-hyp} along the vertices in the unique path from $q$ to $r$ reveals that
    \begin{equation}
        x_q = x_r \prod_{i \in \mathcal P} \frac{v_{c_i}}{v_i - s}
    \end{equation}
    However, it is already known that $x_q=0$ from \rep{root-0}.
    Since for each $i \in \mathcal P$, $v_i - s \neq 0$ (since $r \notin  \mathcal P$),
    and $v_{c_i} \neq 0$ (since only $v_q = 0$ and for no $i \in \mathbf m$ does $q = c_i$),
    if follows that $x_r = 0$ as well.
  From this,~\rep{root-0}, and~\rep{inductive-hyp}, it follows that for all $i \in \mathbf m$, $x_i = 0$.
\qed

\noindent{\bf Proof of Lemma \ref{daniel}:} Let $\tilde{c}_i =\text{column} \{c_{ij_1^i},c_{ij_2^i}, \cdots, c_{ij_{k_i}^i}\}$.
Thus $\tilde{C}_i=\tilde{c}_i\otimes I_{n\times n}$.
By Lemma \ref{G-exists} there are matrices $h_i$ such that $(\sum_{i=1}^m b_ih_i\tilde{c}_i,b_q)$ is controllable with controllability index $m$ for every choice of $q\in \mathbf{m}$.
By definition $\tilde{B}_i=b_i\otimes I_{n\times n}$, $\tilde{C}_i=\tilde{c}_i\otimes I_{n\times n}$ for $i\in \mathbf m$.
Choose $\hat H_i=h_i\otimes I_{n\times n}$.
From this,
\begin{equation*}
  \sum_{i=1}^m \tilde B_i\hat H_i\tilde{C}_i =\left(\sum_{i=1}^m b_ih_i\tilde{c}_i\right) \otimes I_{n\times n}
\end{equation*}
Thus
{\small \begin{align*}
Q \!\triangleq \!&   \matt{\tilde{B}_i \! & \!\left(\!\!\sum_{i=1}^m \!\!\tilde B_i\hat H_i\tilde{C}_i\!\!\right)\!\!\tilde{B}_i  \!& \!\ldots &  \left(\!\!\sum_{i=1}^m\!\! \tilde B_i\hat H_i\tilde{C}_i\!\!\right)^{m-1}\!\!\!\!\tilde{B}_i}  \\ \! =\! &
   \matt{b_i\!\! & \!\left(\!\sum_{i=1}^m\!b_ih_i\tilde{c}_i\!\right)b_i \!& \! \ldots \!\!& \! \left(\!\sum_{i=1}^m \! b_ih_i\tilde{c}_i\! \right)^{m-1\!\!\!}b_i}\!\!\otimes \!\!I_{n\times n}
\end{align*}}
\noindent Since  $(\sum_{i=1}^m b_ih_i\tilde{c}_i,b_q)$ is controllable with controllability index $m$ for every choice of $q\in \mathbf{m}$,
$\text{rank } Q =mn$.
Therefore for each $q\in \mathbf{m}$, $\left(\sum_{i=1}^m \tilde B_i\hat H_i\tilde{C}_i, \tilde B_q\right) $ is a controllable pair with controllability index at most $m$.
On the other hand, note that the matrix $Q$ has exactly $nm$ columns, $m$ is the smallest possible controllability index. Thus for each $q\in \mathbf{m}$, $\left(\sum_{i=1}^m \tilde B_i\hat H_i\tilde{C}_i, \tilde B_q\right) $ is a controllable pair with controllability index $m$.
\qed

\begin{lemma}\label{g}Let   $(A_{n\times n},B_{n\times r})$ be a real-valued controllable matrix pair with
 controllability index $m$.  For each real matrix $M_{n\times n}$, the matrix pair $(M+gA,B)$ is controllable with
  controllability index no greater than
  $m$ for all but at most a finite number of values of the real scalar gain $g$. Moreover if $mr=n$, then $m$ is
   the controllability index of $(M+gA,B)$  for all but a finite number of values of  $g$.\end{lemma}

\noindent{\bf Proof of Lemma \ref{g}:} The assumed properties of
the pair  $(A,B)$ imply that $mr\geq n$ and that there must be a  minor of order $n$ of the matrix
 $\matt{B &AB &\cdots &A^{m-1}B}$ which is nonzero. Let $1,2,\ldots q$ be a labeling of the $n$th order minors of
  $\matt{B &AB &\cdots &A^{m-1}B}$ and suppose that the $k$th such minor is nonzero.
Let $\mu:\R^{n\times n}\oplus \R^{n\times r}\rightarrow \R$ denote that function which assigns to any
matrix pair $(\bar{A}_{n\times n},\bar{B}_{n\times r})$, the  value of
 the $k$th minor of $\matt{\bar{B} &\bar{A}\bar{B} &\cdots &\bar{A}^{m-1}\bar{B}}$.
  Thus $\mu(A,B) \neq 0$ and if $(\bar{A},\bar{B})$ is a matrix pair for which $\mu(\bar{A},\bar{B})\neq 0$,
  then $(\bar{A},\bar{B})$ is a controllable pair with controllabilty index no greater than  $m$.

Since $\mu(A,B)\neq 0$, it must be true that $\mu(gA,gB)\neq 0$ provided $g\neq 0. $   Note that
$\mu(\lambda M+ A,B)$ is a polynomial in the scalar variable $\lambda $.
 Since $\mu(\lambda M+ A,B)|_{\lambda = 0}\neq 0$, $\mu(\lambda M+ A,B)$ is not the zero polynomial.
  It follows that there are at most a finite number of values of $\lambda $ for which $\mu(\lambda M+ A,B)$
  vanishes and $\lambda =0 $ is not one of them. Let $g$ be any number for which $\mu(\frac{1}{g}M+A,B)\neq 0$.
  Then $\mu(M+gA,gB)\neq 0$ and since $g\neq 0$, $\mu(M+gA,B)\neq 0$. Therefore $(M+gA,B)$ is a controllable pair
   with controllability index no greater than $m$.

   Let   $m_g$ denote the controllability index of $(M+gA,B)$; then $m_gr\geq n$.
  Suppose that $mr = n$. It follows that $m_gr\geq mr$ and thus that  $m_g\geq m$.
  But for all but at most a finite set of values of $g$,
  $m_g\leq m$.Therefore $m_g=m$ for all but at most a finite set of values of $g$.
 \qed

\noindent{\bf Proof of Lemma \ref{lem2}:}
As an immediate consequence of Lemma \ref{g} it is clear that for any $K_i,\;i\in\mathbf{m}$ and
 for all but a finite number of values of $g$, the matrix pair $\left (\tilde {A}  +
 \sum_{i=1}^m \tilde{B}_i(K_i\hat{C}_i+g\hat{H}_{i}\tilde{C}_{i}),\tilde{B}_q\right )$
is controllable with controllability index $m$ for every $q\in\mathbf{m}$.
Setting $H_i=g\hat{H}_i$ thus gives the desired result. $\qed$

\noindent{\bf Proof of Proposition \ref{maint}:}
The existence of the $H_i$ which makes the matrix pair $\left (\tilde {A}  +
 \sum_{i=1}^m \tilde{B}_i(K_i\hat{C}_i+H_{i}\tilde{C}_{i}),\tilde{B}_q\right )$
 controllable  for every $q\in\mathbf{m}$ implies that
 all of the complementary subsystems of \rep{anal} are complete \{cf.  Theorem 1 of \cite{corfmat}\}.
From this, the joint controllability and joint observability of \eqref{anal}, it now follows from Corollary 1 of \cite{corfmat}  that
there exist matrices $K_i$ and $H_{ij}$ for which \rep{anal} is  controllable and observable for any value of $q\in\mathbf{m}$.
The matrix pair $\left (\tilde {A}  +
 \sum_{i=1}^m \tilde{B}_i(K_i\hat{C}_i+H_{i}\tilde{C}_{i}),\tilde{B}_q\right )$  also has  controllability index $m$.
Moreover the set of $K_i$ and $H_{ij}$ for which this is true is the complement of a proper algebraic set in the linear space of all such matrices so almost
 any choice for such matrices will have the required properties. $\qed $

\section{Distributed Set-Point Control}
This aim of this section is to explain how the ideas discussed in the preceding sections can be used to solve
 the ``distributed set-point control problem.'' This problem will be formulated assuming that each agent $i$ senses a scalar output
  $y_i = c_ix$ with the  goal of adjusting $y_i$  to  a  prescribed number
    $r_i$ which is agent $i$'s desired {\em set-point} value. The
  {\em distributed set-point control problem} is then to develop a distributed feedback control system for a process modelled
by the multi-channel system \rep{mcs} which, when applied will enable each and every agent
 to independently adjust its output to any desired set-point value.

To construct such a control system, each agent $i$ will make use of  integrator dynamics of the form
\eq{\dot{w}_i = y_i-r_i,\;\;i\in\mathbf{m}\label{int}}
where $r_i$ is the desired \{constant\} value to which $y_i$ is to be set. The combination of these
 integrator equations plus the multi-channel system described by \rep{mcs}, is thus  a system of the form
\eq{\dot{\tilde{x}} = \tilde{A}\tilde{x} + \sum_{i=1}^m\tilde{B}_iu_i-\tilde r\hspace{0.3in} w_i = \tilde{c}_i\tilde{x}, \;\;i\in\mathbf{m}\label{b1}}
where $\tilde{x} = \text{column}\{x,w_1,w_2,\ldots, w_m\}$,
$$\tilde{A} = \matt{A &0\\C & 0},\;\;\;\;\;\;\tilde{B}_i = \matt{B_i\cr 0},\;i\in\mathbf{m},\;\;\;\;\;\tilde{r}= \matt{0\cr r}$$
$C= \text{column}\{c_1,c_2,\ldots, c_m\}_{m\times n}$, $r = \text{column}\{r_1,r_2,\ldots,r_m\}$, and
$\tilde{c}_i = \matt{0 & v'_i}$,   $v_i$ being the $ith$ unit vector in $\R^{ m}$.
Thus \rep{b1} is an $n+m$ dimensional,  $m$ channel system with measurable outputs $w_i,\;i\in\mathbf{m}$, control inputs $u_i,\;
i\in\mathbf{m}$, and constant exogenous input $\tilde{r}$. Note that {\em any} linear constant feedback control, distributed or not,
which stabilizes this system, will enable each agent to attain its desired set-point value. The reason for this is simple.
  First note that any such control will  bound the state of the resulting closed loop system and cause the state
   to tend to a constant limit as $t\rightarrow\infty$. Therefore, since each $w_i$ is a state variable, each must tend
    to a finite limit. Similarly each $y_i$ must also tend to a finite limit.
     In view of \rep{int}, the only way this can happen is if each $y_i$ tends to agent $i$'s desired set-point value $r_i$.

To solve the distributed set-point control problem it is enough to devise a distributed  controller which stabilizes \rep{b1}.
This can be accomplished using the ideas discussed earlier in this paper provided \rep{b1} is both
jointly controllable by the $u_i$ and jointly observable through the $w_i$. According to  Hautus's lemma \cite{hautus},
 the condition for joint observability is
that
$$\rank \matt{sI-\tilde{A}\\ \tilde{C}} = n+m$$
 for all complex number $s$ where $\tilde{C} = \text{column} \{\tilde{c}_1 ,\tilde{c}_2,\ldots,\tilde{c}_m\}$.
In other words what is required is that
 \eq{\rank \matt{sI-A  &0\\ -C &sI\\0 &I} = n+m\label{ober}}
 But $(C,A)$ is an observable pair because \rep{mcs} is a jointly observable
  system. From this,  the Hautus  condition, and the structure of the matrix pencil appearing in \rep{ober} it is
  clear that the required rank condition is
  satisfied and thus that \rep{b1} is a jointly observable system.

  To establish joint controllability of \rep{b1}, it is enough to show that
$\rank \matt{sI-\tilde{A} &\tilde{B}} = n+m$ for all complex number $s$
 where $\tilde{B}  =\matt{\tilde{B}_1 &\tilde{B}_2 &\cdots &\tilde{B}_m}$.
In other words what is required  is that
{\small \eq{\rank \matt{sI-A \! & \! 0 \! & \!B_1 \!&\!B_2 \!&\!\cdots \! &\!B_m\\ -C\! &\!sI\!&\! 0\! &\! 0\!&\! \cdots \!&\!0}=n+m\label{thy}}} But since \rep{mcs} is a jointly controllable system, $\rank \matt{sI-A & B} = n$
 for all $s$, where $B= \matt{B_1 &B_2&\cdots &B_m}$. Thus \rep{thy} holds for
  all $s\neq 0$. For $s=0$, \rep{thy} will also hold provided
\eq{\rank\matt{A & B\\C &0} = n+m\label{gnd}}
In other words,  \rep{gnd} is the condition for \rep{b1} to be jointly controllable and
 thus stabilizable with distributed control.

It is possible to give a simple interpretation of  condition \rep{gnd} for the case when each $B_i$ is a single column. In this case
the transfer matrix $C(sI-A)^{-1}B$ is square and condition \rep{gnd} is equivalent to the requirement that its determinant
has no zeros at $s=0$ \{cf, \cite{asm:struc}\}. Note that if the transfer matrix were nonsingular but had a zero at $s=0$, this would lead to a pole zero
 cancellation at zero because of the integrators.

Suppose condition \rep{gnd} is satisfied. The process of constructing an observer-based  distributed control to stabilize \rep{b1} is as follows.
The first step would be to construct an observer-based  distributed control to stabilize  the  reference signal free system
\eq{\dot{\tilde{x}} = \tilde{A}\tilde{x} + \sum_{i=1}^m\tilde{B}_iu_i\hspace{0.3in} w_i = \tilde{c}_i\tilde{x}, \;\;i\in\mathbf{m}\label{b2}}
using the technique discussed earlier in the paper. This would result in a feedback control system of the form
\begin{eqnarray*}u_i &=&F_ix_i,\;i\in\mathbf{m}\\
\dot{x}_i &=& (\tilde{A}+k_i\tilde{c}_i)x_i -k_iw_i +
 \sum_{j\in\scr{N}_i}H_{ij}(x_i-x_j)
 \\ &\ &
 +\delta_{iq}\bar{C}z + \sum_{j=1}^m\tilde{B}_jF_jx_i  \label{q1}\\
\dot{z} & = & \bar{A}z  +\bar{k}\tilde{c}_qx_q -\bar{k} w_q + \sum_{j\in\scr{N}_q}\bar{H}_{j}(x_q-x_j) \label{q2}  \end{eqnarray*}
Application of this control system to \rep{b1} would stabilize \rep{b1} and thus
 provide a solution to the distributed set-point control problem despite the fact that the signals $x_i$ would not be asymptotically correct
 estimates of $\tilde{x}$.

\section{Transmission Delays}\label{ddlay}
An important issue not discussed so far is the effect of network transmission
delays on the concepts discussed in this paper. While this topic is involved enough to warrant
 treatment in a separate paper, to illustrate one way to deal with delays
  a brief discussion will be given here using a specific example.
 Suppose the system to
  be controlled
  is a three channel,  jointly controllable, jointly observable
   discrete-time system of the form
  \eq{x(t+1) = Ax(t)+\sum_{i=1}^3B_iu_i(t),\;\;\;\;\;y_i(t) = C_ix(t),
   \label{dic}}
   for $i\in\{1,2,3\},\;t\in[0,1,2,\ldots)$.

\noindent Suppose in addition that $\mathbb{N}$ is the directed
graph determined by neighbor sets $\scr{N}_1 =\{1,2\}$, $\scr{N}_2 =
\{1,2,3\}$ , and $\scr{N}_3 = \{2,3\}$. Assume that $u_i =
F_ix_i,\;i\in\{1,2,3\}$ where $x_i$ is agent $i$'s estimate of $x$
and like before, that state feedback matrices $F_i$ have been chosen
so that the matrix $A+B_1F_1+B_2F_2+B_3F_3$ has desired properties
such as discrete-time stability. Like before, the  goal is to
construct a  discrete-time distributed observer with state estimates
$x_i,\;i\in \{1,2,3\}$ so that for each $i$,  the estimation error
$e_i(t) = x_i(t)-x(t)$, converges to zero in discrete time,
exponentially fast at a prescribed rate. Suppose that at time $t$,
agent $1$ receives the  delayed state $x_2(t-2)$ from neighbor $2$
rather than $x_2(t)$, that agent $2$ receives  $x_3(t)$ and the
delayed states $x_1(t-1)$ from its neighbors $3$ and $1$
respectively, and that agent $3$ receives the delayed state
$x_2(t-2)$ from neighbor $2$. One way to  obtain a distributed
observer which delivers the desired behavior, is to use the $2$-unit
delayed states $x_i(t-2),\;\;i\in\{1,2,3\}$ in the local estimators.
In particular, suppose that for $i\in\{1,2,3\}$, the update equation
for agent $i$'s estimator is

{\small
\begin{eqnarray*}x_i(t+1)\!\!\!\!\! & = &\!\!\!\!\!(A+K_iC_i)x_i(t)
- K_iy_i(t) +\left ( \sum_{j=1}^3B_jF_j\right )x_i(t)
\\ &\ & +\sum_{i=1}^3H_{ij}(x_i(t-2)-x_j(t-2))  +\delta_{iq}\bar{C}z(t)
x_i\\
z(t+1)\!\! \!\!\!&=& \!\!\!\!\!\bar{A}z(t) + \bar{K}C_qe_q(t) +\!\!\! \sum_{j\in\scr{N}_q}\!\!\!\bar{H}_j(x_q(t-2)-x_j(t-2))\end{eqnarray*}}
for some fixed  $q\in\{1,2,3\}$. Here $\delta_{iq}$ is the Kronecker delta, just like before.

After picking $q\in\{1,2,3\}$, the first step  in defining the  the $K_i$ and $H_{ij}$ is
 to define the ``lifted''
 error states
  set $e_{ik} = e_i(t-k),\;\;i\in\{1,2,3\}, \;\;k\in\{1,2\}$.  In this case, the open-loop error system is
\begin{eqnarray}e_i(t+1) \!\!& = &\!\!(A+\tilde{K}_iC_i)e_i(t) +\sum_{i=1}^3\tilde{H}_{ij}(e_{i2}(t)-e_{j2}(t)) \nonumber \\ &\ & +
 \sum_{j=1}^3B_jF_j(e_i(t)-e_j(t))
+\delta_{iq}\tilde{u}_q(t)\nonumber\\
 e_{i1}(t+1) \!\!&=& \!\!e_i(t)\nonumber\\
 e_{i2}(t+1) \!\!& =& \!\!e_{i1}(t),\nonumber\end{eqnarray}
 $i\in\{1,2,3\}$, together with the  output
$$\tilde{y}_q(t) = \matt{C_qe_q(t)\\
e_{q2}(t)-e_{j_12}(t)\\
e_{q2}(t)-e_{j_22}(t)\\
\vdots \\
e_{q2}(t)-e_{j_{k_q}2}(t)}$$
where $\{j_1,j_2,\ldots, j_{k_q}\} =\scr{N}_q$.
Using exactly the same techniques that were used earlier in this paper, it can be checked that
 the $\tilde{K}_i$ and
$\tilde{H}_{ij}$ can  be chosen to make this system controllable through input $\tilde{u}_q$ and observable through
output $\tilde{y}_q$. Having so chosen these matrices, the remaining steps  to be taken to define
the $K_i, H_{ij},\bar{A}, \bar{K} $
 and $\bar{H}_j$ are exactly the same  as before in the delay-free case. What results is a distributed observer with the
 requisite properties.

In summary, the key change needed to account for the transmission delays is to use
 the terms $x_i(t-2) - x_j(t-2), j\in\scr{N}_i$ in the update equations for the $x_i$, rather than terms of the form
 $x_i(t) - x_j(t), j\in\scr{N}_i$. The significance of the $2$-unit delay used here is that it is
 also is the maximum of all  transmission delays
  across the network.  This idea generalizes and will be elaborated on   in another paper.

 \section{Concluding Remarks}

 The discussion in \S \ref{ddlay} explains by example one way to deal with network transmission delays. Contrary to intuition, the example illustrates that   transmission delays do {\em not} preclude achieving arbitrarily fast estimation error convergence
 with a suitably defined observer. What delays certainly will effect is estimator ``performance''  in the face of sensor noise and disturbances.
 Quantifying this is a major open problem for future research.

 While the algorithms discussed in this paper can be implemented in a distributed manner, all require ``centralized designs.''  Centralized designs are implicitly assumed  in essentially all decentralized control and distributed control research including, for example, the work in \cite{wangdavison,corfmat}. In our view it is highly unlikely, if not impossible, to avoid centralized designs unless very restrictive assumptions
are added to the problem formulations.
Of course there are some distributed algorithms such as those studied in \cite{TAC.17.2,lineareqn} which do not  call for centralized designs; but these are not feedback control algorithms.

Algorithms based on centralized designs tend to be ``fragile'' in that they will typically fail if there is a single break in the network or perhaps a single component failure. It is thus of interest to try to find
 new algorithms for controlling a multi-channel linear system which require
 ``less'' centralized  designs than assumed  in this paper. Some of the other approaches to observer design cited at the beginning of this paper may prove useful in this regard.

\bibliographystyle{unsrt}
\bibliography{my,steve}

\end{document}